\documentclass[useAMS]{mn2e}
\usepackage{graphicx}

\title[Decade time-scale modulation of low mass X-ray binaries]{Decade
  time-scale modulation of low mass X-ray binaries}
\author[Durant et al.]{Martin Durant$^1$\thanks{E-mail:
durant@iac.es}, Remon Cornelisse$^1$, Ron Remillard$^2$ and Alan Levine$^2$\\
$^{1}$ Instituto de Astrof\'isica de Canarias, Calle V\'ia L\'actea
  s/n, 38205 La Laguna, Tenerife, Spain\\
$^{2}$ Kavli Institute for Astrophysics and Space Research, MIT, 77
  Massachusetts Ave., Cambridge, MA 02139, USA}
\begin{document}

\date{March 2009}

\pagerange{\pageref{firstpage}--\pageref{lastpage}} \pubyear{2009}

\maketitle

\label{firstpage}
\begin{abstract}
Regular observations by the All Sky Monitor aboard the Rossi X-ray
Timing Explorer satellite have yielded well-sampled light-curves with
a time baseline of over ten years. We find that up to eight of the
sixteen brightest persistent low mass X-ray binaries show significant,
possible sinusoidal, 
variations with periods of order ten years. We speculate on its
possible origin and prevalence in the population of low mass X-ray
binaries and we find the presence of a third object in the system, or long-period
variability intrinsic to the donor star, as being attractive origins
for the X-ray flux modulation we detect. For some of the
objects in which we do not detect a signal, there is substantial
short-term variation which may hide modest modulation on long time-scales. Decade time-scale modulations may thus be even more common.

\end{abstract}


\section{Introduction}
Low mass X-ray binaries (LMXBs) are systems in which a compact object
accretes matter from a late-type ordinary donor star (main sequence or
evolved), which is filling its Roche lobe.
They can either be transient or persistent X-ray sources. The former, 
 as the name suggests,  exhibit very large changes in timing,
 spectral, and luminosity properties. Transient outbursts are observed
 with a wide range of 
time-scales from hours to $\geq$10\,yr.  
Persistent LMXBs, although they undergo out-bursts and flaring episodes, 
have not been seen to change in luminosity by more than a factor of a
few. They have not, by definition, been seen to
enter {\em quiescence}, a very low luminosity state.
They typically contain a neutron star as the accreting compact
object.
The persistent LMXBs
include the so-called Z-sources, Accretion Disc Corona (ADC) sources,
and Atoll Sources, where the naming scheme reflects their 
behaviour on a X-ray hardness/luminosity or colour-colour
diagram. Charles and Coe (2006) review observational properties. 

The {\em Rossi} X-ray Timing Explorer (RXTE, Bradt et al., 1993;
Jahoda et al. 1996) is a satellite
dedicated to time-domain investigation of X-ray sources. It includes
three instruments: the Proportional Counter Array (PCA) for pointed
observations in the soft-medium energy band (2--30\,keV), the
High-energy Timing Experiment for simultaneous high-energy
observations (20--200\,keV) and the All-Sky Monitor (ASM, Levine et
al., 1996). The ASM comprises three shadow cameras, and scans 20\% to 80\% of the sky
every 90\,min in the energy range 2–-10\,keV. It produces light-curves for
all sources of interest (with a spatial resolution of 3--15$^\prime$),
has been operational since 1996, and now provides light-curves for 564
sources. (Note: all raw data is archived, and light-curves can be
generated for a given position a-posteriori.)

Accreting binary systems manifest a wide range of timing signals:
kHz oscillations and ms bursts/pulsations (possibly from the neutron star
surface, Lamb \& Miller, 2001), Hz and sub-Hz quasi-periodic oscillations (QPOs) from the inner parts of the accretion 
disc, and mHz ``slow'' QPOs perhaps due to feedback processes within
the accretion disc (e.g., Bozzo et al., 2009). The time-scale
of variability produced by changes in the disc will be related to
local Keplerian frequencies, and may also be related to the convection
time-scale or viscous in-fall time-scale. Added to this are
the periodicities associated with the orbital system: the orbital
period and the precession  in the disc, which may be the cause of {\em
  super-orbital} such as for Her X-1 (Scott \& Leahy, 1999). Clearly, timing of
X-ray binaries is a  rich source of information on the system dynamics.

Whereas high-frequency, i.e., Hz and kHz, QPOs are extremely well-studied in the literature (e.g., van der Klis, 2004),
long time-scale variability is not, since  its study requires homogeneous observations
with good sampling over time base-lines beyond most observing
campaigns. The RXTE/ASM archive provides light-curves of the brightest
X-ray sources on the sky in a homogeneous way, and with regular observations.
The sample we choose are the persistent neutron star X-ray binaries
with typical count rates above $\sim$5\,cts\,s$^{-1}$ in the ASM; this
is the same sample as in Reig et al. (2003), less Cyg X-1 (a black
hole) but plus Ser X-1. In an investigation of the
long time-scale variability of a number of bright sources in X-ray monitoring data,
Reig et al. (2003) plotted power density spectra down to frequencies
of the order 10$^{-7}$\,Hz (0.3\,yr), and interpreted the spectra in
terms of red noise.

Wen et al. (2006) were the latest to perform a systematic search for
periodicities (or quasi-periodicities) in RXTE/ASM data (see also
Levine \& Corbet, 2006). They discovered a number of significant peaks
in Lomb-Scargle periodograms (Scargle, 1989), but only in a few cases did they
report evidence for periods over 100\,days. The longest period reported was that of 4U
1820$-$303, and there were also two further objects  with power spectrum peaks
classified as quasi-periodic.

For details of the objects considered in this work,  see the X-ray
binary catalogues of Ritter \& Kolb (2003) and Liu et
al. (2007) and their respective updates, and many references
therein. For details on high-frequency timing characteristics, see van
der Klis (2004).
We summarise brief details of our sample in Table \ref{objects}.

\begin{table*}
\begin{center}
 \centering
  \caption{Details of objects in our sample}\label{objects}
  \begin{tabular}{lcccccc}
  \hline
   Name     & Type & Bursts & kHz QPOs & Orbital Period & Comments \\
 \hline
GX 9+9 & Atoll & & yes& 4.2\,hr & The only object for which a multi-year periodicity has\\
& & & & &already been found and discussed (Harris et al., 2009)\\
GX 9+1& Atoll & & yes & \\
GX 354$-$0& Atoll& &yes & &Month time-scale variability in Kong et al. (1998)\\
GX 3+1& Atoll&yes & yes& &Makishima et al. (1983) give some evidence of
multi-\\
&&&&&year variability\\
Ser X-1& Atoll& yes& yes& \\
4U 1735$-$444& Atoll& yes&yes & 4.7\,hr\\
Sco X-1& Z-source& & yes& 18.9\,hr&  \\
4U 1636$-$536& Atoll&yes &yes & 3.8\,hr& Shih et al. (2005) found
$\sim$40\,day periodicity\\
& & & & & (see also Belloni et al. 2007)\\
GX 340+0& Z-source& & yes\\
4U 1705$-$44 & Atoll &  yes &no &1.3\,hr? & possible 223\,d on/off
cycle (Priedhorsky \& Holt, 1987)\\
4U 1820$-$30 & Atoll& yes& yes&0.19\,hr &176\,d cycle, \~Simon (2003)\\
Cyg X-2 & Z-source& ? & yes& 236.2\,hr&39\,d period (and strong harmonics, Boyd \& Smale. 2004)\\
GX 13+1 & Atoll& ? & & &24\,d period, possible disc warp (Corbet 2003)\\
GX 17+2 & Z-source & yes & no & & \\
GX 349+2 & Z-source &  & yes\\
GX 5$-$1 & Z-source &  & yes\\
\hline\\
\end{tabular}\\
\end{center}
\end{table*}

In Section 2 we analyse light-curves from the ASM aboard RXTE. We find
that six show clear modulation with periods of the order
10 yr, and that two more show variability that suggests periodic or
quasi-periodic behaviour.  We fit these with a simple  
function, and thereby obtain estimates of periods and amplitudes. In
Section 3 we discuss the possible causes of periodicities that could
produce such variations, and the prevalence of 
modulation in the population of persistent X-ray binaries.

\section{Analysis and results}
\begin{figure*}
\includegraphics[width=\hsize]{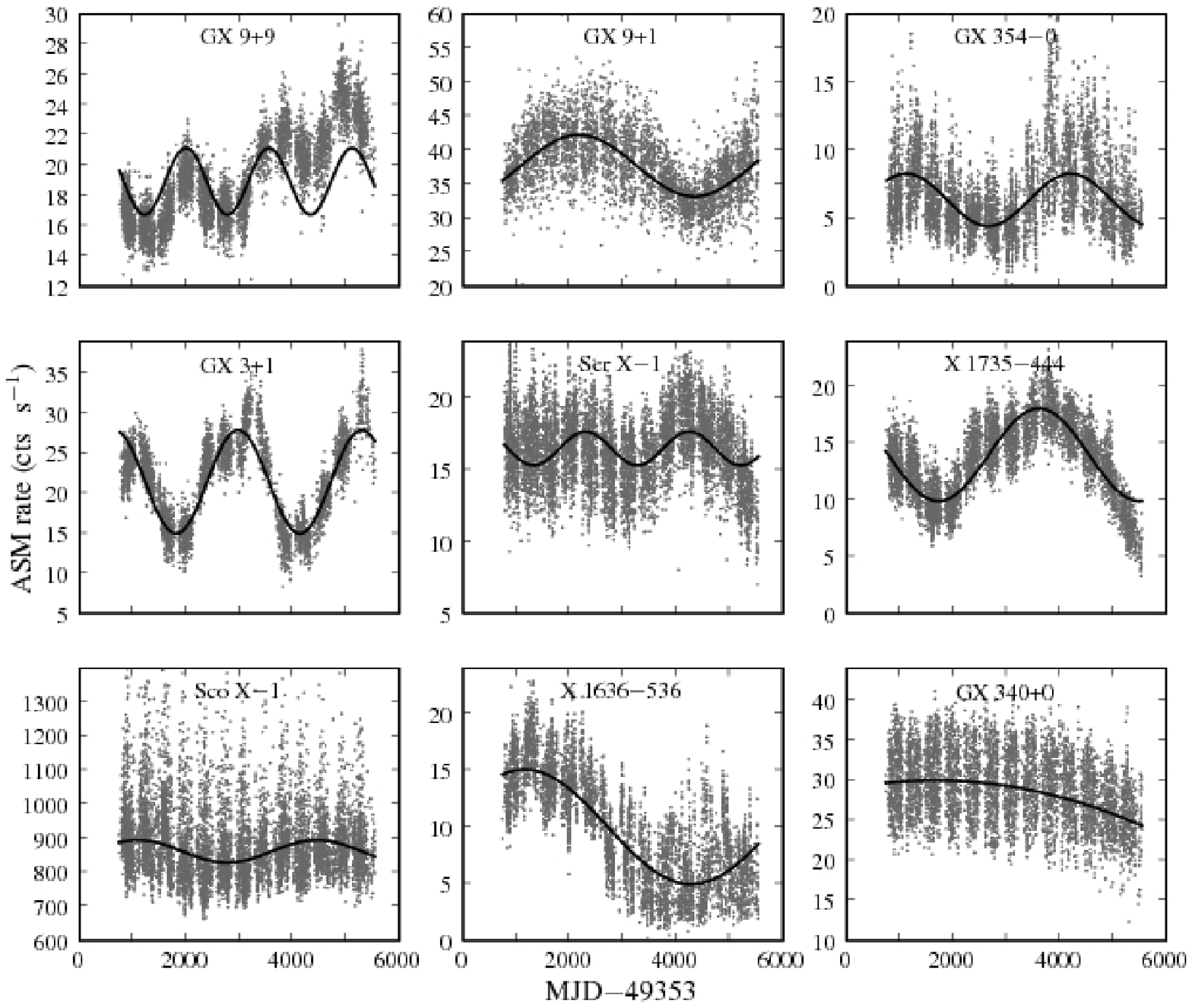}
\caption{ASM light-curves of the sample of X-ray binaries for which we
  fitted a sine plus constant model. The X-axes
  span $\sim$16\,yr in every case. Over-plotted are the best-fit
  sinusoids for each data-set, with parameters as given in
  Table~\ref{pars}. The vertical bands in the data are due to periods
  when observations were not possible, see text. The
  line is not a feasible fit for GX 340+0 (see Table \ref{pars}), but
  we show it as our limiting case.}\label{lc} 
\end{figure*}

We retrieved ASM data, as ten-observation averages across all the
available time, from the {\em ASM Weather Map} web-page\footnote{{\tt
    http://heasarc.gsfc.nasa.gov/xte\_weather/}} of the sources in Reig et al.'s
(2003) neutron star binary sample plus Ser X-1. The data for GX 9+1
was not available from that web page 
at the time of writing; instead we retrieved one-day averages from the
ASM Light Curves Overview 
page\footnote{{\tt
    http://xte.mit.edu/ASM\_lc.html}}. Nine light-curves are plotted
in Figure~\ref{lc}. They are all well-sampled, with 3900--7600 points
per light-curve. The vertical banding is due to those times when the
source was too close to the sun, and therefore no observations were
possible. They occur, therefore, once per year, with a typical gap size of order 1-2months, depending on the source position and brightness. We note from the outset, that although the light-curve of GX
340+0 is shown here, the fitted period is much longer than the
observation time base-line; we only show it as an example of a {\em
  long-term trend} (see below). 

\begin{figure*}
\includegraphics[width=\hsize]{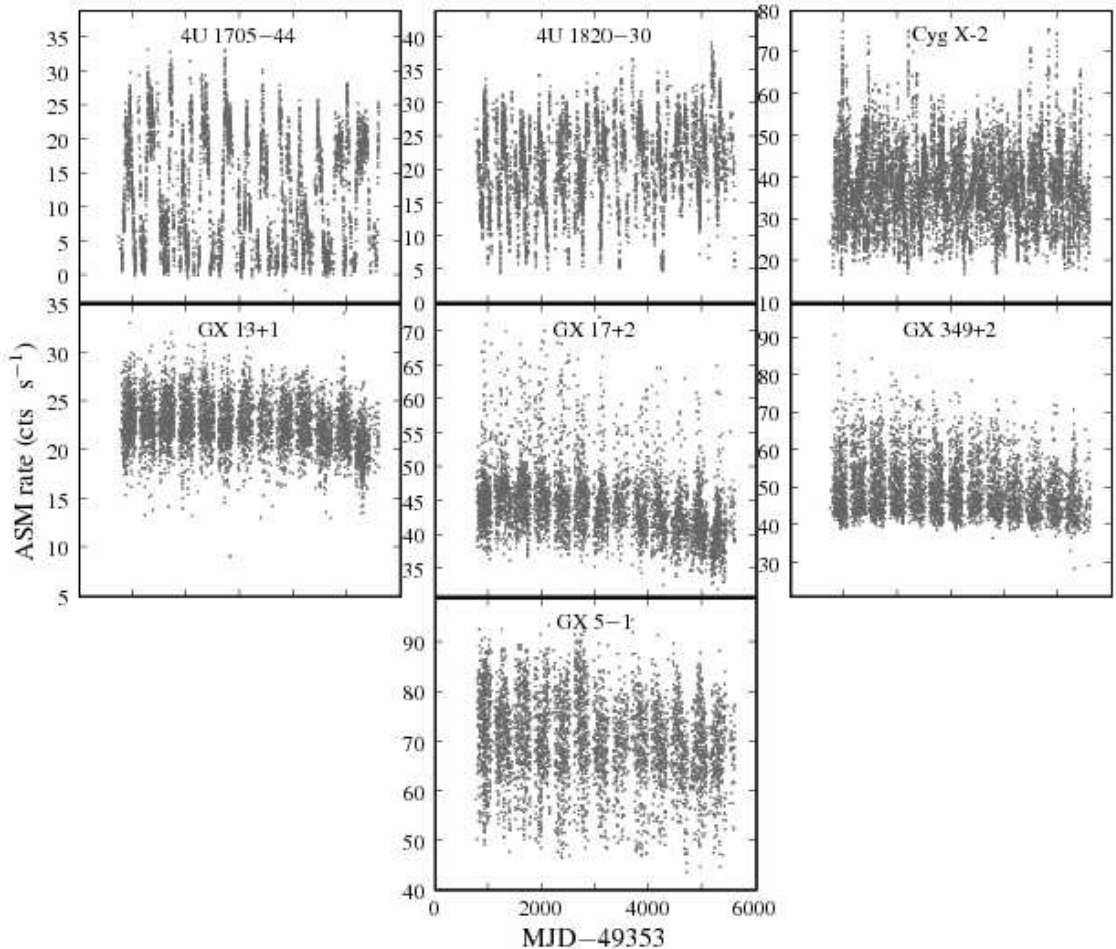}
\caption{ASM light-curves of the sample of X-ray binaries for which we
  did not fit a periodic model. }\label{lc2} 
\end{figure*}

\begin{figure*}
\includegraphics[width=\hsize]{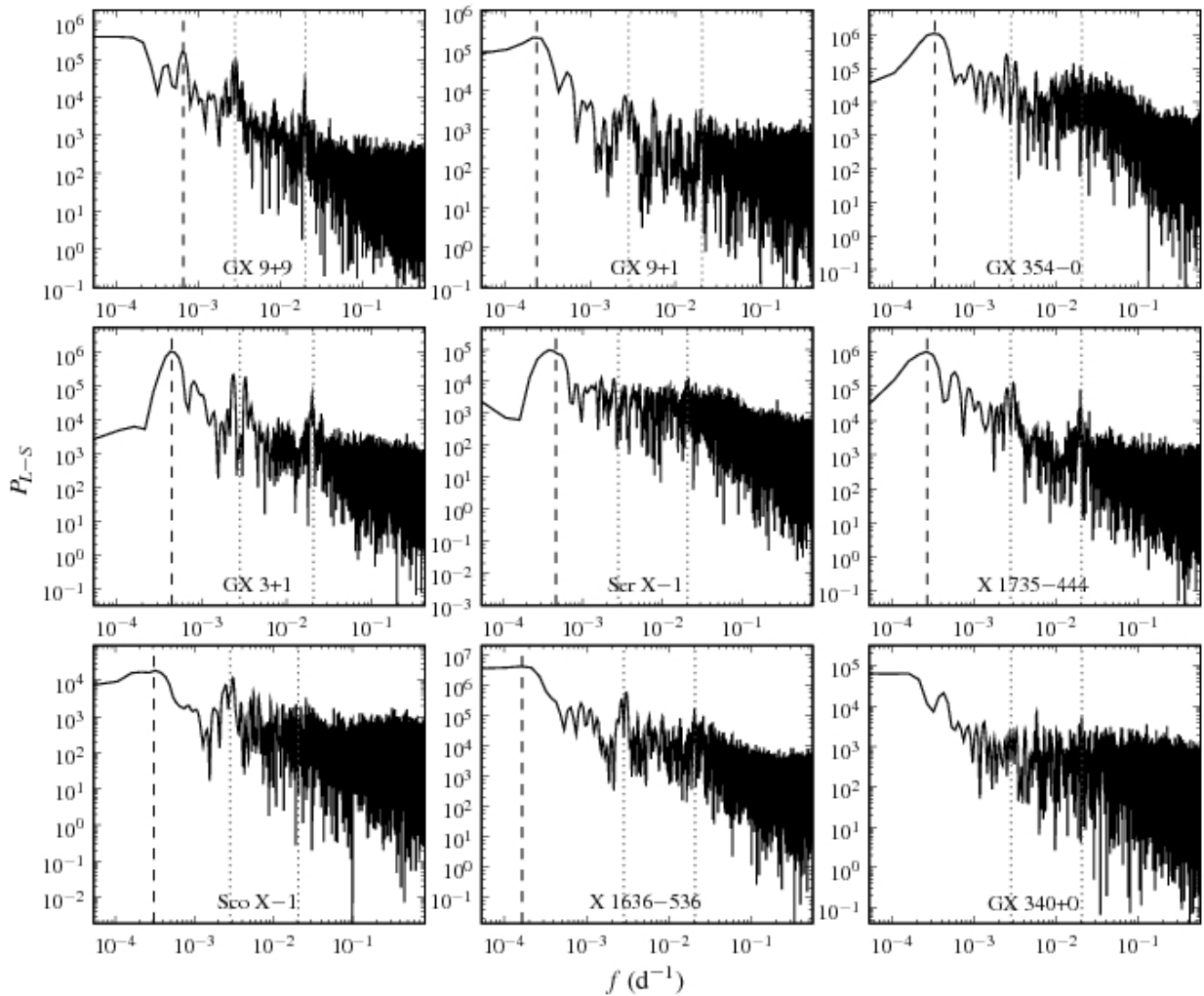}
\caption{Lomb-Scargle power spectra of the light-curves of the sources
for which we fitted sine functions in Figure \ref{lc}. Vertical dotted lines
indicate periods of one year and 50 days, which appear starkly on some
of the plots. The vertical scale is standard RMS$^2$/Hz. Also shown by
the black dashed line, are the periods found by fitting and shown in
Table \ref{pars}.}\label{yes} 
\end{figure*}
\begin{figure*}
\includegraphics[width=\hsize]{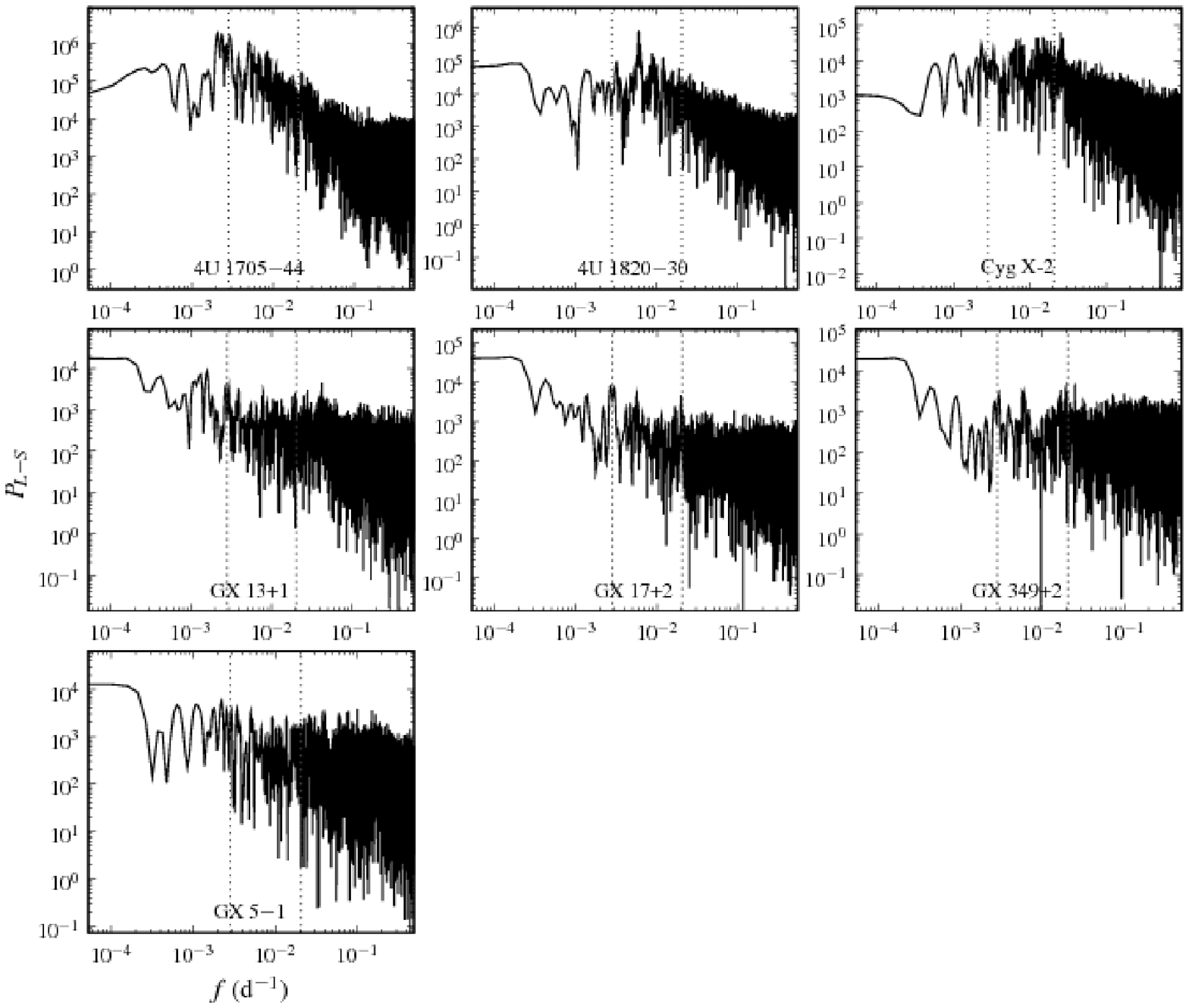}
\caption{Lomb-Scarge power spectra as in Figure \ref{yes}, but for the
  sources where there was no clear strong signal to fit with a sine
  function. Certain known periodicities/QPOs are evident: 223\,d for
  4U 1705$-$44, 173\,d for 4U 1820$-$30, 78\,d for Cyg X-2 (second
  harmonic of the known super-orbital period), and 24\,d for GX
  13+1. }\label{no}  
\end{figure*}

A sine-like modulation is apparent in a number of the light curves.  Although
clearly no simple model, such as a sinusoid, completely describes the
variability in each source (or even necessarily the form of the
slow modulation), a simple approach
gives useful numeric estimates of amplitudes and periods.  Given that
only three or fewer cycles are seen, we note that the determination of
a period and amplitude does not imply the presence of a true
periodicity; this is discussed in more detail below. 

To each data-set,
we fitted a simple function of the form
\begin{equation}
F= C + A \times \sin(2\pi(d - Z)/P) \label{eq}
\end{equation}
giving the flux $F$ at day $d$. $C$ is a constant, $A$ is the
amplitude of the sinusoid, $Z$ is the time of zero phase and $P$ is
the period. The given uncertainties in each measurement
are rather small, typically less than 1\,cts\,s$^{-1}$. Although
we use these uncertainties in our least-squares fits in the
calculation of weights, they 
clearly do not reflect the scatter seen in the measured values,
which may be due to additional activity in the source,
e.g., short-term flaring. In order, therefore, to derive
uncertainties in the fit parameters, we have scaled the given
uncertainties such that the 
best-fit model gives  $\chi^2/dof = 1$. Note that the uncertainties
on each ASM measurement are rather similar throughout a
single light-curve, and unweighted fits would yield essentially the same parameters 
and uncertainties thereon as obtained in our weighted fits.

\begin{table*}
\begin{center}
 \centering
  \caption{Categorisation and best-fit parameters (where available) for the data, as
    given by Equation \ref{eq} and plotted in Figure \ref{lc}. $Z$ was
  chosen to be near to time $d=0$. Numbers in parentheses are
  1-sigma uncertainties on the last digit(s), assuming the fit is
  reasonable (see text).}\label{pars}
  \begin{tabular}{lccccccc}
  \hline
   Name     & Sine-like & Period $P$& Amplitude $A$ & Constant $C$
   & Phase Zero $Z$ & $\sigma(F)/\sigma(F-model)$ & P$_{L-S}$~$ ^2$\\
 & modulation& (d) & (cts\,s$^{-1}$) & (cts\,s$^{-1}$) & (MJD$-$49353) \\
 \hline
GX 9+9 & yes & 1558(8) &      2.20(5) &     18.95(3)
&49(10)&1.17&$4.2\times 10^5$\\
GX 9+1& yes &4340(30) &       4.50(8)&      37.74(6)&1090(20)&1.26&
$2.3\times 10^5$\\
GX 354$-$0&yes&3085(30) &       1.92(5)&       6.36(4)&340(30)&1.15&
$1.2\times 10^6$\\
GX 3+1&yes&2321(7) &       6.44(6)&      21.44(5)&72(9)&1.93&
$1.2\times 10^6$\\
Ser X-1&possible&1945(15) &       1.17(4)&
16.52(3)&$-$135(13)&1.06& $1.0\times 10^5$\\
4U 1735$-$444&yes&3770(14) &       4.08(4)&
13.99(3)&$-$1105(6)&1.72& $1.1\times 10^6$\\
Sco X-1&possible&3364(60)& 33.1(2)& 861.5(12)& 246(60)&1.02& $2.0\times 10^4$\\
4U 1636$-$536&possible&6254(80)  &     5.05(5)  &
10.02(6)&$-$411(60)&1.66& $4.5\times 10^6$\\
\hline
GX 340+0$^1$&trend&68600& 90& $-$60 & $-$15500&1.08&$6.8\times10^4$\\
4U 1705$-$44 & short-term$^3$ & & & & & & $2.9\times10^5$\\
4U 1820$-$30 & short-term$^3$ & & & & & & $8.7\times10^4$\\
Cyg X-2 & short-term$^3$ & & & & & & $1.6\times10^4$\\
GX 13+1 & trend & & & & & & $1.8\times10^4$\\
GX 17+2 & trend & & & & & & $4.6\times10^4$\\
GX 349+2 &trend & & & & & & $2.2\times10^4$\\
GX 5$-$1 &no & & & & & & $1.3\times10^4$ \\
\hline\\
\end{tabular}\\
$^1$: This fit is highly uncertain, and sample numbers are only
  indicative. $C<0$ is clearly unphysical. \\
$^2$: highest Lomb-Scargle power for periods $P>1$\,yr.\\
$^3$: too much variability on short time-scales to be able to reliably determine if there is any slow modulation.
\end{center}
\end{table*}

 By eye, the fits to the light-curves generally appear to be
reasonable. Certainly, a constant plus a sinusoid is preferred
over a constant (see  Table
\ref{pars}). However, a sinusoid may not be better than
other periodic functions or polynomials. Nevertheless, it grossly
characterises the variability, and the periods give a good
idea of the time-scales involved. 

We list all of the objects analysed in Table
2 together with information, where available, on the results of the fits.
The fit curves are shown, together with the original data
in Figure \ref{lc} for the cases where we believe a model fit to be
at least a first-order description of the trend in the data.  We
also include in Figure 1 the light curve of one binary, GX 340+0, that
serves as an example of the light-curves which are designated 
''trend'' in Table 2, wherein the fit produces a period
much longer than the observation base-line, and therefore
we cannot even claim to have detected a sine-like modulation. The
remaining light-curves are shown in Figure \ref{lc2}. The light-curves
of some of these sources are dominated by very strong variability on
time scales of 40 to 200 days. 

The source 4U 1636-536 is an interesting special case. Since MJD 51353
it has shown state transitions between low/hard and high/soft states
with the total flux changing by a factor $\sim$10, and having a peak
value near the typical flux before this behaviour commenced (Farrell
et al. 2009).  This
makes the fit in Table \ref{pars} more troublesome, yet it still
appears to first order to fit the data points as presented.

Alternative methods are also generally available to assess
periodicities, such as phase-dispersion minimisation
(PDM; see Schwarzenberg-Czerny, 1997, for a thorough description of the
relevant statistics). In the case we have here, however, with data
sets covering less than 
two periods, such methods do not work reliably, as the statistics
are dominated by the number of samples in each bin and
therefore the number of bins chosen; they become equivalent
to coarse re-binning and sine fitting. The power spectrum
method below is roughly equivalent to the method of fitting a sine function,
but it provides interesting information at all frequencies.

In Figures \ref{yes} and \ref{no} we also show the Lomb-Scargle power
spectra for all the sources, those with putative periods and those
without, respectively (following Scargle, 1989).  For each source where we find
a periodic solution, this corresponds to the lowest-frequency
isolated peak in the power-spectra, with the exception of
GX 9+9, where there is an additional high shoulder at the lowest
frequencies (see Harris 
et al., 2009). The diagram for GX
340+0 shows a long-term trend, where the peak is still rising at the low-frequency
cut-off. This is similar to the other light-curves labeled as
long-term trends in Table \ref{pars}.
 Note the spurious peaks at $\sim$~1\,yr and $\sim$50\,d in Figures \ref{yes} and \ref{no}. These peaks are artificial, due to the gaps
 in the data when the source is near the sun: once per year a source is unobservable for a typical $\sim$50\,d duration (the exact size and
 frequency of the shorter-period peak depends on how close the source
 passes to the sun). Another possible origin for the $\sim$50\,d peak is the precession period of the spacecraft, which affect the number of observations possible during each orbit (e.g., Corbet, 2003).
This leaves little sample space left to allow us
 to measure the red-noise, and thus independently show the
 significance 
of the putative periods. We have included the value of the peak power
in each power spectrum in Table \ref{pars} for reference.

As an interesting aside, we note that the spurious peaks in  Figures
\ref{yes} and \ref{no} are commonly double-peaked about the
contaminating frequency. This kind of behaviour was noted by Farrell
et al. (2005) at higher frequencies, and ascribed by them to power
leakage in the sampling. It may be that similar processes are at work
here. 

We have also investigated
the light-curves of the three separate ASM energy
bands (nominally 1.5--3.0, 3.0--5.0 and 5.0--12\,keV).
We find that the slow modulations are similar in the three bands and,
therefore, there do not appear to be any 
changes in the ratios of fluxes between these bands that are
correlated to any large extent with the overall flux. 

Having found a number of sinusoidal modulations with amplitudes of
$\sim$10\%--50\%, with various periods, for different objects, it
seems very difficult to ascribe this effect to some calibration or
other instrumental effect. Indeed, the performance of the ASM is
checked regularly against the observed emission of the Crab, and
corrected for any long-term change in sensitivity (Levine et al.,
1996). Note that for some objects in the ASM catalogue,
the intensity is stable over many years. See also the
discussion in Harris et al. (2009) concerning the fluxes
measured in the ASM and the PCA instruments of RXTE - certainly for
the case of GX 9+9, the light-curve modulations appear to be real.

\section{Discussion}

We have found that the light curves of up to eight of 16 bright LMXBs appear
to have sine-like modulations that may be characterized by
periodicities with periods in the range 1500--6250\,d (4--17\,yr) and
relative amplitudes ($A/C$) of 7--50\%. A simple function consisting
of a constant plus a sinusoidal function of time provides a first 
order description of the data, but there is also shorter time-scale
variability and scatter and even some long-term trends that are not
fit by this simple model. With periods similar to the duration of the
total observing window, it is not possible to say if the modulation
is strictly periodic, quasi-periodic or chaotic. That the sines
fit so closely the forms of some of the light-curves, however, suggests
the presence of some kind of periodicity or quasi-periodicity.

The modulation amplitudes above are rather large compared to the X-ray
modulations seen due to orbital motion. Excepting eclipsing or
nearly-eclipsing 
systems (called ``dippers''), the orbital amplitude is rarely more
than a couple of percent (Liu et al., 2007), and indeed some orbital
periods have 
been found from the larger-amplitude optical modulation. Super-orbital
periods can be stronger, however: for instance, the amplitude of the 176 day semi-periodic modulation of 4U 1820$-$03 is about a factor of 2 (i.e.,
$A/C\sim1/3$,  Zdziarski et al. 2007), in the same range as our
numbers. In a few cases, super-orbital modulations are large enough to
be described, in part, as turn on/off events, e.g., as in the case of
the 35 day super-orbital period of Her X-1 (Scott \& Leahy, 1999). As
already mentioned above, 
short-term variations of luminosity by factors $>$2 are not uncommon,
including some of the sources considered here (see Figure \ref{lc} and
Table \ref{pars}). 

It is is interesting that it seems to be mostly the Atoll sources which
show the type of modulation considered here. Indeed, they seem to be
more likely to have shorter-term (of order
30\d) modulations also. One might suspect that this was due to large
fractional erratic variability among the Z-sources, but the
light-curves in Figures \ref{lc} and \ref{lc2} do not appear to bear
this out. There may be a clue here to what makes the difference
between Z- and Atoll sources. There is no obvious relation to bursting
or kHz QPO behaviour.

A number of concrete possibilities can be considered as origins
for the periods we detect. Since the X-ray emission is powered by
accretion onto the compact object in each binary, it is natural to
attribute a change in observed flux to a change in accretion rate. In
LMXBs, accretion occurs through the transfer of mass from the donor
star through the L1 point (known as Roche lobe over-flow); if the
un-deformed radius of the star or its magnetic field configuration
change on time-scales of order decades, this can explain the
modulations we see. Long-period variable stars are known (Wood, 1987), and
indeed the sun has a well-studied 11\,yr magnetic configuration
cycle. 

The presence of a third body in a binary system such as a low-mass
star, could result in the presence of long-term
periodicities. If the modulation seen has the same periodicity as the
orbit of this third body, it would need to come close enough in its
orbit to tidally interact with the inner binary system at some point
during its orbit. The size of the inner binary is determined by the
requirement that matter can transfer out of the Roche lobe of the
donor, and the total mass is dominated by the neutron star. Thus, for
the given periods and a typical K-type or smaller donor star, the
orbit of the third body would be of order 100 times the inner binary
separation, and may require some eccentricity to get close
enough to cause tidal interactions; in this case, a smoothly varying
light-curve may be hard to explain.
 With an orbital period of the order days,
however, it can interact with the disc and 
with the binary. The interaction can induce precession within the
inner binary, or possibly directly in the accretion disc though tidal
resonances, causing accretion flow variations that have the
time-scales of beat periods, much longer 
than the third body orbital period (Mazeh \&
Shaham, 1979; Ford et al., 2000). Zdziarski et al. (2007) used such a model
to explain the 176\,day modulation seen in 4U 1820$-$30.

Periodicities due to the disc alone are also possible.  For example,
precession may occur with respect to the orbital plane and the neutron star
spin/magnetic field. The periodicities may be the result of feedback
cycles, such  as commonly invoked to 
explain the much stronger cyclic/episodic behaviour seen in X-ray
transients. In these, the build up of a magnetic field suppresses the
accretion on to the neutron star, but is subsequently suppressed when
the disc has built up enough matter (e.g., Lasota, 2008). 

We cannot envision a scenario in which the observed X-ray modulations
are caused by variable amounts of obscuration, e.g., like that
resulting from a precessing warp in the accretion disc (see Ogilvie \& Dubus,
2001; Clarkson et al. 2003).  Increases in the 
intervening column density of cool or partially ionized gas would
produce stronger variations in the softest energy channel, and this effect is
not observed. Alternatively, if the intervening gas were highly
ionized, then its origin and substantial modulation would have to be
hypothesized without any clear physical motivation.  Partial
obscuration of the X-ray source by an opaque medium is also a
possibility, but it may be difficult to tie such structures to the
accretion disc, given the small angular size of the X-ray source and
the smooth sinusoidal appearances of the X-ray modulation.

To summarise our findings: out of the sixteen sources considered, we
found evidence for sine-like modulation in six, with two more possible cases,
and four with long-term variability on time-scales too long for us to
be able to determine the form of the modulation or if it is
periodic. Only four sources showed no sign of such a signal, and even 
here for three we may speculate that it is present, but hidden under
more prominent short-term variability. Thus it would seem that within
the persistent neutron star LMXB population, multi-year periodic or quasi-periodic
variability is common.  Indeed, some of the systems for which we found
no clear modulation are those for which long super-orbital periods
were already known (Cyg X$-$2, 4U 1820$-$30 and GX 13+1). If the
modulation for 4U 1636$-$536 and GX 354$-$0 are real and
(quasi-)periodic, they would be a new class of variability, since both
systems already have known super-orbital periods.

When the operation of the RXTE ASM terminates some time in the next
few years, it will not be possible
to monitor these light-curves further in exactly the same
way. Hopefully, either the coming MAXI sky monitor on the
International Space Station or the
ASTROSAT mission with its sky monitor or both will provide a
comparable or even better capability.
The Swift and INTEGRAL satellites perform
regular observations of much of the sky; if they operate over long
enough times, their results from somewhat higher 
photon energies may also be able to provide important information
about the modulations of the binary systems studied in this work .

\section*{Acknowledgments}
MD is supported by the Spanish Ministry of Science under grant
AYA20042646; MD and RC receive 
some support from Consolider-Ingeniero 2010 Program grant
CSD2006-00070; RC acknowledges a Ramon y Cajal
fellowship (RYC-2007-01046).
RXTE is operated by NASA.

\bsp

\label{lastpage}


\begin{thebibliography}{99}
\bibitem[]{a}{Belloni, T., Homan, J., Motta, S., Ratti, E., M\'endez, M.,
  2007, MNRAS, 379, 247}
\bibitem[]{bozz}{Bozzo, E., Stella, L., Vietri, M., Ghosh, P., 2009,
  A\&A, 493, 809}
 \bibitem[]{bs}{Boyd, P.~T., \& Smale, A.~P., 2004, ApJ, 612, 1006}
\bibitem[]{xte}{Bradt, H., Rothschild, R., Swank, J., 1993, A\&AS, 97,
  355}
\bibitem[]{cc}{Charles, P. \& Coe, M., 2006, In: Compact stellar X-ray
  sources, eds Lewin \& van der Klis. Cambridge
  Astrophysics Series, No. 39. Cambridge University Press} 
\bibitem[]{clark}{Clarkson, W., Charles, P., Coe, M., Laycock, S.,
  2003, MNRAS, 343, 1213}
\bibitem[]{gxxg}{Corbet, R., 2003, ApJ, 595, 1086}
\bibitem[]{wooohoo}{Farrell, S., Barret, D., Skinner, G., 2009, MNRAS,
  393, 139}
\bibitem[]{leaky}{Farrell, S., O'Neill, P., Sood, R., 2005, PASA, 22,
  267} 
\bibitem[]{ford}{Ford, E., Kozinsky, B., Rasio, F., 2000, ApJ, 535, 385}
\bibitem[]{b} {Harris, R., Levine, A., Durant, M., Kong, A., Charles, P.,
  Shahbaz, T., 2009, ApJ, 696, 1987}
\bibitem[]{jah}{Jahoda, K., Swank, J., Giles, A., Stark, M.,
  Strohmayer, T., Zhang, W., Morgan, E., 1996, SPIE, 2808, 59}
\bibitem[]{kon}{Kong, A., Charles, P., Kuulkers, E., 1998, NewA, 3, 301}
\bibitem[]{bah}{Lamb, F. K., Miller, M., C., 2001, ApJ, 554, 1210}
\bibitem[]{las}{Lasota, J.-P., 2008, NewAR, 51, 752}
\bibitem[]{asm}{Levine, A., Bradt, H., Cui, W., Jernigan, J., Morgan,
  E., Remillard, R., Shirey, R., Smith, D., 1996, ApJ, 469, L33}
\bibitem[]{c}{Levine, A. \& Corbet, R., 2006, ATel 940}
\bibitem[]{cat2}{Liu, Q., van Paradijs, J., van den Heuvel, E.,
  2007, A\&A, 469, 807}
\bibitem[]{maki}{Makishima, K., Mitsuda, K., Inoue, H. et al., 1983,
  ApJ, 267, 310}
\bibitem[]{third}{Mazeh, T. \& Shaham, J., 1979, A\&A, 77, 145	
}
\bibitem[]{og}{Ogilvie, G \& Dubus, G., 2001, MNRAS, 320, 485}
\bibitem[]{prid}{Priedhorsky, W., \& Holt, S., 1987, Space Science
  Reviews, v45 no3, 291}
\bibitem[]{lll}{Priedhorsky, W. \& Terrell, J., 1984, ApJ, 284, L17}
\bibitem[]{reig}{Reig, P., Papadakis, I., Kylafis, N., 2003, A\&A, 398,
  1103}
\bibitem[]{cat}{Ritter, H., Kolb, U., 2003, A\&A, 404, 301}
\bibitem[]{L-S}{Scargle, J., 1989, 343, 874}
\bibitem[]{PMD}{Schwarzenberg-Czerny, A., 1997, ApJ, 489, 941}
\bibitem[]{herx}{Scott, D. \& Leahy, D., 1999, ApJ, 510,974}
\bibitem[]{d}{Shih, I., Bird, A., Charles, P., Cornelisse, R., Tiramani,
  D., 2005, MNRAS, 361, 602}
\bibitem[]{sim}{\~Simon, V., 2003, A\&A, 405, 199}
\bibitem[]{snail}{Smale, A. \& Lochner, J., 1992, ApJ, 395, 582}
\bibitem[]{vdk}{van der Klis, 2004, in Compact stellar X-ray sources, Lewin \& van der Klis (eds), Cambridge University Press, see {\tt astro-ph.10551}}
\bibitem[]{e}{Wen, L., Levine, A., Corbet, R., Bradt, H., 2006, ApJS,
  163, 372}
\bibitem[]{sdv}{Wood, P., 1987, LNP, 274, 250}
\bibitem[]{svd}{Zdziarski, A., Wen, L., Gierliński, M., 2007,
  MNRAS, 377, 1006}

\end{thebibliography}
\end{document}